\documentclass[aps,prd,twocolumn,notitlepage,10pt,
nofootinbib,nobibnotes,superscriptaddress,preprintnumbers]{revtex4-2}

\usepackage[utf8]{inputenc}
\usepackage[T1]{fontenc}

\usepackage{physics}
\usepackage{graphicx}
\usepackage{dcolumn}
\usepackage{bm}
\usepackage{booktabs}
\usepackage{makecell}
\usepackage{siunitx}
\usepackage{subcaption}
\usepackage[hidelinks]{hyperref}

\begin{document}

\title{Strangeness Production in Heavy-Ion Collisions:\\ Color Ropes or Hydrodynamic Evolution?}

\author{Carl B. Rosenkvist}
\email{rosenkvist@itp.uni-frankfurt.de}
\affiliation{Frankfurt Institute for Advanced Studies, Ruth-Moufang-Strasse 1, 60438 Frankfurt am Main, Germany}
\affiliation{Institute for Theoretical Physics, Goethe University, Max-von-Laue-Strasse 1, 60438 Frankfurt am Main, Germany}
\affiliation{GSI Helmholtzzentrum f{\"u}r Schwerionenforschung, Planckstr. 1, 64291 Darmstadt, Germany}

\author{Hannah Elfner}
\affiliation{GSI Helmholtzzentrum f{\"u}r Schwerionenforschung, Planckstr. 1, 64291 Darmstadt, Germany}
\affiliation{Institute for Theoretical Physics, Goethe University, Max-von-Laue-Strasse 1, 60438 Frankfurt am Main, Germany}
\affiliation{Helmholtz Research Academy Hesse for FAIR (HFHF), GSI Helmholtz Center, Campus Frankfurt, Max-von-Laue-Stra{\ss}e 12, 60438 Frankfurt am Main, Germany}

\date{\today}

\begin{abstract}
We investigate strangeness production and transverse dynamics in heavy-ion collisions at $\sqrt{s_{\mathrm{NN}}}\approx 2.5\text{–}20~\mathrm{GeV}$ using the transport approach SMASH (Simulating Many Accelerated Strongly-interacting Hadrons), its extension with rope hadronization, and the SMASH+vHLLE hybrid approach. Results from the Pythia-based heavy-ion model Angantyr, with and without rope hadronization, are included for comparison. We study midrapidity particle yields and average transverse masses as functions of the number of wounded nucleons, as well as their energy dependence.

For the $K^+/\pi^+$ ratio, SMASH+vHLLE overpredicts strangeness production at low energies but describes the higher-energy behavior reasonably well. SMASH+Ropes reproduces the ratio up to $\sqrt{s_{\mathrm{NN}}}\sim\SI{10}{GeV}$ but does not capture the turnover at higher energies. In contrast, the transverse-mass observables favor the hybrid approach, while the non-thermal models considered here do not generate sufficient collective transverse expansion.

These results show that strangeness enhancement alone does not uniquely distinguish microscopic string interactions from a locally equilibrated medium. Simultaneously constraining strangeness production and transverse dynamics is therefore essential for disentangling thermal and non-thermal mechanisms in heavy-ion collisions.
\end{abstract}

\maketitle

\section{Introduction}
One of the primary goals of heavy-ion collisions is to explore the quantum chromodynamics (QCD) phase diagram, which describes the behavior of strongly interacting matter under extreme conditions of temperature and density. Both theoretical and experimental studies suggest the existence of a phase known as the quark-gluon plasma (QGP), where quarks and gluons are nearly deconfined \cite{Harris_1996,Gyulassy:2004zy,Shuryak:2004cy,CMS:2011iwn,PHOBOS:2004zne,STAR:2005gfr,ALICE:2010suc}.

Strangeness provides a particularly useful probe in this context because it is conserved under the strong interaction and must be produced during the collision. In a purely hadronic phase, strange particles can only be created as fully formed hadrons, which is strongly suppressed due to their relatively large masses and the need to assemble a complete hadronic state at once. By contrast, a deconfined or partially deconfined medium can produce strangeness more efficiently, since it allows the creation of individual $s\bar{s}$ pairs without immediately forming a bound hadron. These pairs can later hadronize into strange and multi-strange particles, greatly reducing the suppressive effect present in hadronic reactions. As a result, rapid $s\bar{s}$  production from gluon interactions is expected to enhance strange-hadron yields, especially multi-strange baryons and strange antibaryons, in QGP-forming systems~\cite{PhysRevLett.48.1066,Rafelski:2011ek,Letessier:2022fax}.

Measurements by the ALICE collaboration have extended this strangeness enhancement to small systems, observing enhanced strange particle production in high-multiplicity proton–proton (pp) collisions \cite{ALICE:2016fzo}. While such observations raise the question of whether QGP-like signatures can emerge in small systems, the formation of a long-lived, macroscopic QGP phase in pp collisions is generally not expected due to the small system size and short lifetime. In this regime, collective behavior may instead arise from microscopic, non-thermal mechanisms acting at the level of strings and partons. Indeed, Pythia, which models hadron production via interacting Lund strings, is able to reproduce the observed multiplicity dependence of strange-hadron production in pp collisions when string interactions such as color reconnection and rope hadronization are included~\cite{Bierlich:2014xba,Hushnud:2023mgy}.

This naturally raises the question of whether the necessary building blocks for QGP-like signatures already exist in elementary nucleon–nucleon collisions and can be developed further in larger systems, or if a full QGP phase is essential to describe the experimental data. This idea is explored in the Angantyr~\cite{Angantyr} framework, which extends the same string-interaction picture successful in pp collisions to heavy-ion systems by stacking parton-level nucleon–nucleon events and allowing their strings to interact, thereby generating collective effects without invoking an explicit hydrodynamic phase~\cite{Bierlich:2022ned,Bierlich:2022oja,Chakraborty:2021nde,Lonnblad:2023stc,Chakraborty:2022nui}.

From a complementary perspective, transport models provide a well-established framework to study the microscopic, non-equilibrium dynamics of heavy-ion collisions, particularly at low to intermediate collision energies where full thermalization may not be achieved. In this regime, transport approaches offer a natural setting to investigate how string-based mechanisms compete with, mimic, or eventually necessitate hydrodynamic descriptions of collectivity.
Several transport models commonly used to describe low- to intermediate-energy heavy-ion collisions, such as SMASH~\cite{SMASH:2016zqf}, UrQMD~\cite{URQMD}, and PHSD~\cite{Cassing:2009vt}, employ the Lund string model to represent the initial state.

UrQMD and SMASH have traditionally been used in hybrid approaches, where an initial transport stage is followed by a hydrodynamic evolution once local thermalization is assumed, to capture collective behavior~\cite{Petersen:2009zi,Schafer:2021csj}. The hydrodynamic phase encodes pressure-driven collective expansion and near-equilibrium particle production. Within this framework, strangeness enhancement arises mainly from the high energy densities and thermalized medium described by hydrodynamics, rather than from microscopic non-equilibrium dynamics in the transport phase. For example, the UrQMD hybrid approach has been successfully used to describe NA49 and NA57 data in Ref.~\cite{Petersen:2009zi}.

At the same time, string-interaction mechanisms leading to enhanced strangeness production are not new in transport theory. For instance, UrQMD explored modifications of the strangeness suppression factor of strings through an increased string tension~\cite{Soff:1999et}, while RQMD investigated strangeness enhancement via string fusion into color ropes in Pb–Pb collisions at $E_{\rm lab}=$158A GeV~\cite{Sorge:1995dp}.

We focus on the collision-energy range $\sqrt{s_{\rm NN}}\approx 2.5$–$20$~GeV, which spans the transition from resonance-dominated to string-dominated dynamics and covers a region in which the onset and properties of a hot and dense medium are of particular interest. Analogous to studies of small systems at high energies, investigating this intermediate-energy regime provides an opportunity to determine which microscopic and macroscopic mechanisms dominate the collision dynamics.

SMASH is known to underproduce strangeness in heavy-ion collisions. To investigate whether non-thermal mechanisms can alleviate this, we implement a rope hadronization scheme in SMASH, following~\cite{Bierlich:2014xba} and drawing additional inspiration from string percolation~\cite{Bautista:2019mts}. We refer to this extension as SMASH+Ropes, which provides a string-interaction alternative to the hybrid approaches typically used with SMASH and enables a direct comparison between string-interaction and hydrodynamic modeling within the same framework, thereby reducing model uncertainties.

Since SMASH+Ropes is based on Pythia’s rope hadronization framework, we compare our results to Angantyr, which incorporates this framework directly~\cite{Bierlich:2022ned}. Although Angantyr is not typically applied at such low beam energies, Ref.~\cite{Abdel-Waged:2022ahk} shows good agreement with experimental data at $E_{\rm lab}=40A$ GeV, supporting its use as a reference in this regime.

In parallel, we study strangeness production within the SMASH+vHLLE hybrid model~\cite{Schafer:2021csj}, which combines SMASH with a 3+1D viscous hydrodynamic evolution (vHLLE)~\cite{Karpenko:2013wva}. In this approach, SMASH provides the initial conditions and the hadronic afterburner, while the hydrodynamic stage describes the hot and dense fireball. The transition between hydrodynamics and hadronic transport is handled via the Cooper–Frye particlization procedure. This setup offers a description of an approximately locally equilibrated medium, modeled by viscous hydrodynamics and including controlled deviations from local thermal equilibrium, providing a complementary perspective on how collective dynamics influence strangeness production.

The comparison between SMASH+Ropes and SMASH+vHLLE allows us to directly investigate whether the observed strangeness enhancement requires a locally equilibrated fluid or can emerge from non-thermal color-rope dynamics. If SMASH+Ropes reproduces the strange-hadron yields and patterns obtained with SMASH+vHLLE, this would suggest that these features do not uniquely require a hydrodynamic medium. Conversely, differences between the two approaches can help identify observables that are sensitive to the presence of a hydrodynamic phase.

Section~\ref{sec:models} provides an overview of the different model configurations considered in this work. The baseline models are described in Sec.~\ref{sec:baseline}. Section~\ref{sec:ropes} describes rope hadronization in Pythia/Angantyr and its implementation in SMASH. The hybrid approach is described in Sec.~\ref{sec:hybrid}. Section~\ref{sec:results} presents the results for strangeness production and transverse dynamics as functions of the number of wounded nucleons and collision energy, followed by their discussion in Sec.~\ref{sec:discussion}. Finally, Sec.~\ref{sec:sum} summarizes our conclusions and outlook. The appendices provide details on the determination of the number of wounded nucleons and the software versions used in this work.

\section{Overview of models}\label{sec:models}

In this work, we compare several theoretical frameworks commonly used to simulate heavy-ion collisions. Each framework is based on different assumptions about the underlying microscopic dynamics, equilibration mechanisms, and the origin of collective behavior. The specific software versions used in this study are listed in Appendix~\ref{app:software_versions}.

\textbf{SMASH}~\cite{weil2025smash} is a hadronic transport approach in which hadrons are evolved microscopically through binary scatterings, resonance dynamics, and string excitation at higher energies, without invoking local thermalization.

\textbf{Pythia/Angantyr}~\cite{bierlich2022comprehensiveguidephysicsusage,Angantyr} provides the baseline for string-fragmentation approaches. In Pythia, the Lund string-excitation and fragmentation model is employed, while Angantyr extends this framework to heavy-ion collisions by stacking nucleon--nucleon sub-collisions using Glauber geometry and color-fluctuation effects. Since SMASH employs Pythia for string fragmentation and hard scatterings, this provides a natural point of comparison.

\textbf{Pythia/Angantyr + Ropes}~\cite{Bierlich:2014xba,Bierlich:2017sxk,Bierlich:2022ned} introduces non-thermal collectivity by allowing overlapping color strings to combine coherently into color ropes with enhanced string tension. This mechanism increases the probability of producing strange quarks and baryons.

\textbf{SMASH+Ropes} is an extension developed in this work. It follows the rope-hadronization approach of Pythia/Angantyr but implements it within a space--time transport framework. Strings propagate prior to fragmentation, such that rope formation is determined dynamically from their local density, leading to a time-dependent effective tension and a distinct mechanism of non-thermal collectivity.

\textbf{SMASH+vHLLE}~\cite{Schafer:2021csj} is a hybrid approach in which the early, dense stage of the collision is described using 3+1D viscous hydrodynamics, while the dilute pre- and post-hydrodynamic stages are treated within SMASH.

Together, these frameworks provide a controlled comparison between non-thermal, string-based collectivity and collective dynamics arising from hydrodynamic evolution. In the following, we first describe the baseline approaches Pythia/Angantyr and SMASH, followed by the rope-hadronization mechanism and the hybrid approach in detail.

\section{Baseline Models}\label{sec:baseline}

\subsection{Pythia/Angantyr}\label{sec:pythia}

Pythia is a general-purpose event generator for hadronic collisions at high beam energies, combining perturbative and phenomenological descriptions across different momentum scales. Here, we briefly summarize the aspects relevant for this work, following Ref.~\cite{bierlich2022comprehensiveguidephysicsusage}.

An important component is the multiparton interaction (MPI) framework, in which the rise of the non-diffractive cross section with energy is interpreted as the opening of phase space for multiple semi-perturbative parton--parton scatterings within a single event~\cite{Sjostrand:1987su}. 

The outgoing partons from hard and semi-hard interactions are evolved down to the confinement scale using initial- and final-state parton showers based on DGLAP evolution. During this evolution, color connections are tracked in the large-$N_c$ limit, where each gluon is represented as a color--anticolor pair. At the confinement scale, the linear part of the QCD potential is modeled as a relativistic string with constant energy density (the Lund string model). This string breaks via the tunneling production of quark--antiquark pairs, terminating the color flux tube and producing the hadrons observed experimentally. This process is referred to as string fragmentation.

The transverse momentum distribution of a quark--antiquark pair produced in string breaking is given by
\begin{equation}\label{eq:tunneling_string}
\frac{1}{\kappa} \frac{d\mathcal{P}}{dp_\perp}
\propto
\exp\left( -\frac{\pi m_\perp^{2}}{\kappa} \right),
\end{equation}
where $\kappa$ is the string tension, representing the energy density in the color flux tube. In practice, Eq.~\eqref{eq:tunneling_string} leads to a Gaussian transverse-momentum distribution with width
\begin{equation}
\sigma_\perp^2 = \frac{\kappa}{\pi}.
\end{equation}

From Eq.~\eqref{eq:tunneling_string}, the relative probability for strange-quark production is
\begin{equation}\label{eq:strange_supp_string}
\rho \equiv \frac{\mathcal{P}_s}{\mathcal{P}_{u,d}}
=
\exp\left( -\pi \frac{m_s^2 - m_u^2}{\kappa} \right)\,.
\end{equation}
Using typical values $\kappa = 0.2~\mathrm{GeV}^2$, $m_u \approx 0.3~\mathrm{GeV}$, and $m_s \approx 0.5~\mathrm{GeV}$ yields a suppression factor of approximately $0.08$, which is significantly smaller than required by experimental data. For this reason, the strangeness-suppression parameter $\rho$ is typically treated as a tunable parameter.

The Angantyr model~\cite{Angantyr} extends the Pythia framework from proton–proton to nucleus–nucleus collisions using a Monte Carlo Glauber model incorporating subnucleonic fluctuations. For a given nuclear configuration, the Glauber calculation determines the wounded nucleons and the corresponding nucleon–nucleon ($NN$) sub-collisions, distinguishing between different types of interactions. Individual nucleons can participate in multiple sub-collisions, which are classified as primary or secondary interactions.

Each $NN$ sub-collision is classified as either diffractive or non-diffractive. Non-diffractive interactions involve color exchange and lead to substantial particle production, while diffractive interactions preserve color neutrality and typically produce fewer particles, often accompanied by large rapidity gaps.

In Angantyr, primary and secondary interactions are distinguished. Secondary non-diffractive excitations of nucleons that have already interacted are generated using Pythia’s single-diffraction excitation machinery, with kinematics aligned to the direction of the newly wounded nucleon. This procedure allows the construction of a complete heavy-ion event from an ensemble of individually modeled sub-collisions. Hadronization of the sub-collisions is performed using Pythia’s standard string-fragmentation model.

\subsection{SMASH}
Simulating Many Accelerated Strongly-interacting Hadrons (SMASH) is a relativistic hadronic transport model that includes all well-established hadrons with masses up to about $2~\mathrm{GeV}$ as active degrees of freedom. The dynamical evolution of the system is governed by the relativistic Boltzmann equation,
\begin{equation}
p^\mu \partial_\mu f_i(x,p) + m_i F^\alpha \partial_\alpha^p f_i(x,p) = C_\text{coll}^i ,
\end{equation}
where $f_i(x,p)$ denotes the single-particle distribution function for species $i$. The collision term $C_\text{coll}^i$ accounts for binary scatterings, resonance formation, and decays, while the force term $F^\alpha$ represents mean-field potentials. In the calculations presented here, mean-field interactions are neglected. Collisions are implemented using a geometric criterion: two particles interact if their transverse distance becomes smaller than an interaction radius derived from the total cross section.

At low beam energies, particle production is described by resonance excitation and decay, while above $\sqrt{s} \approx 4~\mathrm{GeV}$ soft string excitation provides the dominant contribution.

The soft-string module includes single-diffractive, double-diffractive, and non-diffractive excitations to strings. In non-diffractive events, each incoming hadron exchanges a valence constituent with the other beam, resulting in the formation of two color strings stretched between a quark (anti-diquark) and an antiquark (diquark). In contrast, diffractive processes excite the hadrons without color exchange.

At higher energies, the description transitions to a hard string-excitation regime. In this regime, string production and fragmentation are handled within Pythia’s multiparton interaction (MPI) framework.

All strings produced in SMASH, independent of their production mechanism, are fragmented using Pythia (Sec.~\ref{sec:pythia}).

\section{Rope Hadronization}\label{sec:ropes}

In high-density environments, multiple color strings may overlap, and their color fields can interact coherently. Such overlapping flux tubes can combine into a higher-dimensional SU(3) representation, known as a \emph{color rope}. This idea, first proposed in the 1980s~\cite{Biro:1984cf,Bialas:1984ye}, provides a non-thermal mechanism for enhanced strangeness and baryon production in hadronic and nuclear collisions.

A key theoretical ingredient is the Casimir-scaling hypothesis~\cite{Ambjorn:1984dp}, supported by lattice-QCD calculations~\cite{Bali:2000un}, which states that the energy density (string tension) of a color flux tube scales with the quadratic Casimir operator of its SU(3) representation. As a consequence, overlapping strings give rise to stronger color fields, increasing the probability of producing heavier quark pairs—particularly strange quarks—as encoded in Eqs.~\eqref{eq:tunneling_string} and \eqref{eq:strange_supp_string}.

The most widely used quantitative realization of this idea is the rope hadronization framework developed by Bierlich \textit{et al.}~\cite{Bierlich:2014xba}. This model is already implemented in Pythia/Angantyr and serves as the theoretical basis for the rope extension introduced for SMASH in this work.

\subsection{Pythia/Angantyr + Ropes}\label{sec:pythia_rope}

In Pythia's rope model~\cite{Bierlich:2014xba}, overlapping strings are combined into an SU(3) multiplet built from $p$ coherent triplets and $q$ coherent antitriplets, forming a representation ${p,q}$ with multiplicity
\begin{equation}\label{eq:rope_multiplicity}
N = \frac{1}{2}(p+1)(q+1)(p+q+2).
\end{equation}

Adding a triplet to ${p,q}$ can produce ${p+1,q}$, ${p-1,q+1}$, or ${p,q-1}$, with probabilities determined by Eq.~\eqref{eq:rope_multiplicity}, generating a random walk in SU(3) space as additional strings overlap.

The effective rope tension follows from the quadratic Casimir operator:
\begin{equation}
\frac{\kappa^{\{p,q\}}}{\kappa^{\{1,0\}}} =\frac{C_2(p,q)}{C_2(1,0)} = \frac{1}{4}\bigl(p^2 + q^2 + pq + 3(p+q)\bigr)
\end{equation}
where $\kappa^{\{1,0\}}$ is the fundamental string tension. For string breaking, the relevant transition ${p,q} \to {p-1,q}$ produces an effective tension
\begin{equation}\label{eq:keff_angantyr}
\kappa_\text{eff} = \kappa^{{p,q}} - \kappa^{{p-1,q}} = \frac{2p + q + 2}{4}\kappa.
\end{equation}

An enhancement factor can then be defined as $h = \kappa_\text{eff}/\kappa$, which modifies key hadronization parameters. In particular, the strangeness suppression factor becomes
\begin{equation}
\tilde{\rho} = \rho^{1/h},
\end{equation}
and the width of the transverse momentum distribution in string breakups increases as
\begin{equation}
\tilde{\sigma}_\perp = \sigma_\perp\sqrt{h}.
\end{equation}

These modifications reflect enhanced pair production and larger transverse momentum kicks in stronger color fields. Other hadronization parameters adjusted in Pythia are described in more detail in~\cite{Bierlich:2014xba}.

\subsection{SMASH+Ropes}

SMASH+Ropes is a modification of SMASH in which strings are propagated through spacetime prior to hadronization. This allows for a dynamical determination of the local string density at the time of fragmentation.

An essential aspect of the framework is the interaction of strings with the surrounding medium, as this directly influences the degree of baryon stopping. If strings are treated as non-interacting, they retain most of their longitudinal momentum and propagate along the beam direction without significant rescattering. As a result, they leave the medium before fragmenting, leading to reduced particle production at midrapidity.

To account for secondary interactions, strings are assigned an effective cross section that depends on the flavors of their endpoints. In particular, a string produced in an $NN$ collision is assigned the nucleon--nucleon cross section. This enables further interactions prior to fragmentation, allowing for secondary excitations and increased baryon stopping.

Following the strategy of Angantyr, these excitations are treated as single--diffractive hadron excitations, with invariant masses sampled from
\begin{equation}
P(M^2) \propto \frac{1}{(M^2)^{\alpha}}.
\end{equation}

\begin{figure}
\centering
\includegraphics[width=0.45\textwidth]{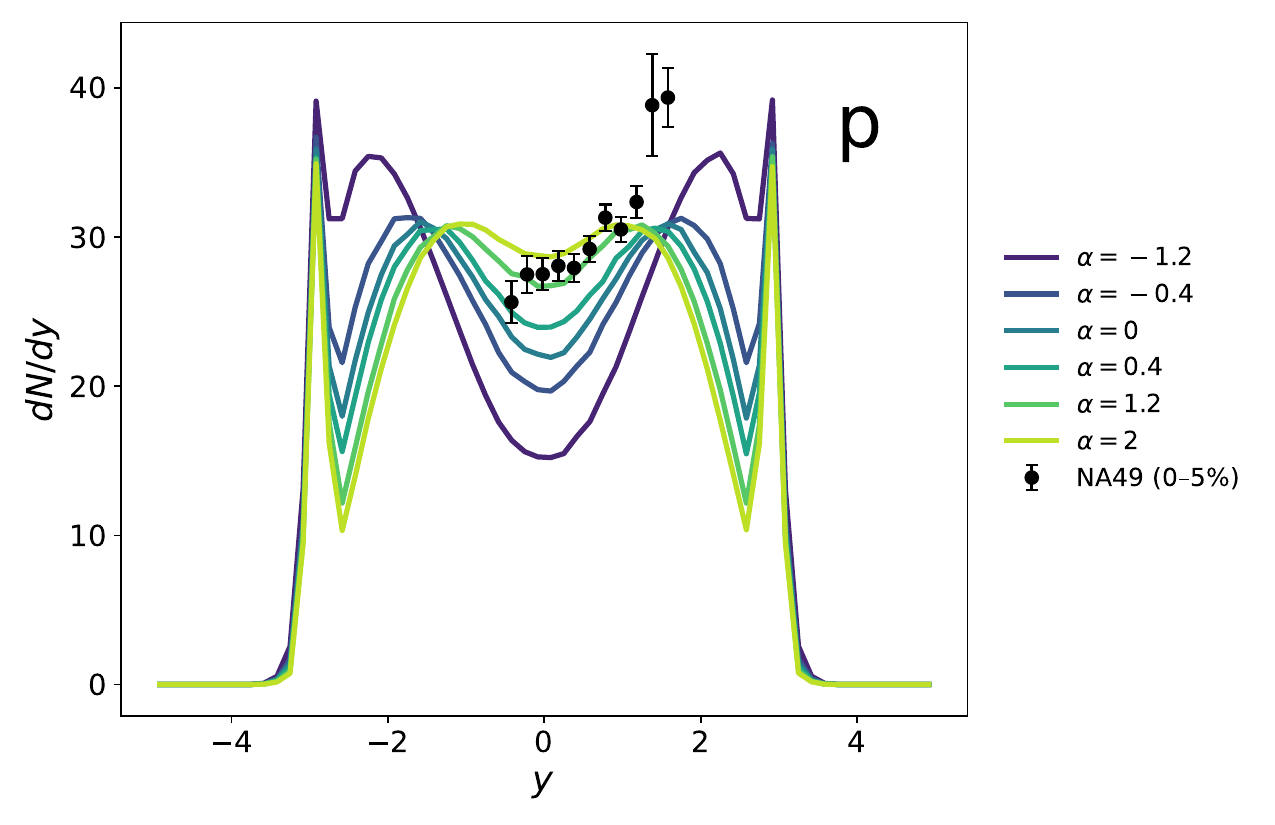}
\caption{Proton rapidity distribution for different values of $\alpha$ in central Pb--Pb collisions at $\sqrt{s_{\mathrm{NN}}}=\SI{17.3}{GeV}$ compared with data from NA49~\cite{Strobele:2009nq}.}
\label{fig:string_stopping}
\end{figure}

Here, the parameter $\alpha$ controls the steepness of the mass spectrum. Smaller values of $\alpha$ favor higher excitation masses and thus larger longitudinal momentum loss, while larger $\alpha$ bias the sampling toward lower masses, closer to the non--interacting limit. In this way, tuning $\alpha$ regulates the average four--momentum transfer from the string to the medium (see Fig.~\ref{fig:string_stopping}). Unlike in Angantyr, where $\alpha$ is tuned to small values, it is treated here as a free parameter controlling the amount of stopping.

With sufficient stopping established, the overlap of strings is evaluated. Inspired by the string-percolation picture, where strings are modeled as overlapping two-dimensional disks~\cite{Bautista:2019mts}, one could in principle formulate the overlap in terms of a two-dimensional transverse geometry, assuming approximately parallel strings. However, this neglects the orientation of strings as well as their extension in the longitudinal direction. Non-parallel string configurations are not considered explicitly here. Bierlich et al.~\cite{Bierlich:2022oja} discuss such configurations in the context of color-rope formation, effectively treating the strings in a frame where they are parallel.

In this work, we instead evaluate the overlap within three-dimensional spatial cells. This should not be interpreted as treating strings as genuine three-dimensional objects with neglected orientation, but rather as an effective measure of the local string density and overlap strength within the SMASH interaction grid.

On the SMASH interaction grid, the probability for a string to participate in rope formation is determined by the local string density. For a cell $c$, it is defined as
\begin{equation}
P_{\mathrm{rope}}(c)
=
\min\left(1,\frac{V_{\mathrm{tot,string}}(c)}{V_c}\right),
\end{equation}
where $V_c$ is the cell volume and
$V_{\mathrm{tot,string}}(c)=\sum_{i\in c}V_i$ is the total effective
volume of the strings contained in the cell. Each string is assigned the
same effective volume, $V_i\equiv V_{\mathrm{string}}$, such that
$P_{\mathrm{rope}}(c)$ corresponds to the local string-volume fraction,
capped at unity.

\begin{figure}
\centering
\includegraphics[width=0.45\textwidth]{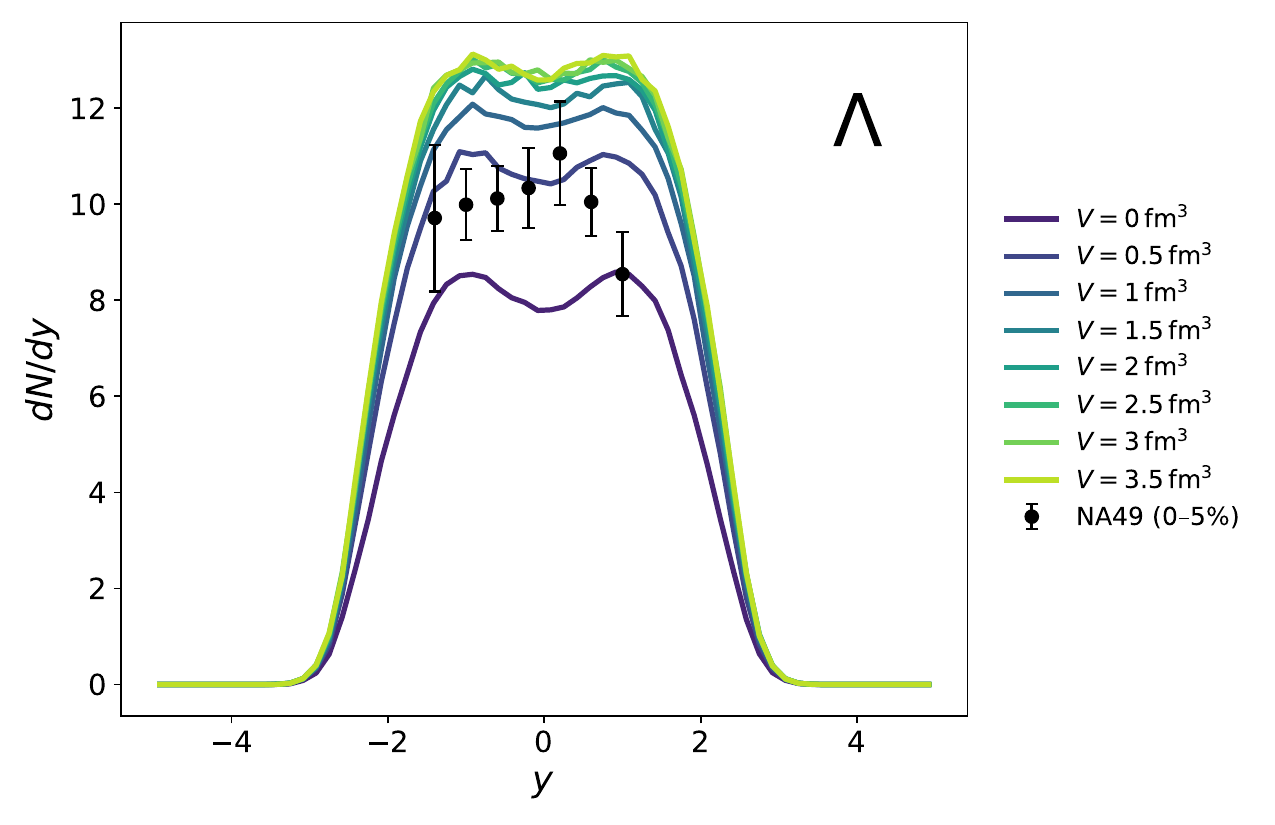}
\caption{$\Lambda$ rapidity distribution for different values of $V_\text{string}$ in central Pb--Pb collisions at $\sqrt{s_\text{NN}}=\SI{17.3}{GeV}$ compared to NA49 data~\cite{NA49:2009flb}.}
\label{fig:string_size_lambda}
\end{figure}

For a cell containing $N$ strings, each string is independently selected
for rope formation with probability $P_{\mathrm{rope}}(c)$. For fixed $N$,
the number of selected strings therefore follows a binomial distribution
with expectation value
\begin{equation}
\langle N_{\mathrm{rope}}\rangle
=
N P_{\mathrm{rope}}(c).
\end{equation}
The parameter $V_{\mathrm{string}}$ thus controls the probability of rope
formation at a given local string density. Its impact on observables is
illustrated in Fig.~\ref{fig:string_size_lambda}.

The rope multiplicity in SMASH+Ropes is determined using the same SU(3) random-walk prescription as in Pythia/Angantyr (Sec.~\ref{sec:pythia_rope}), following Ref.~\cite{Bierlich:2014xba}.

Since each string in SMASH is assigned a finite formation time, multiple strings may fragment near-simultaneously within a given time step. Larger rope segments can therefore break up at once, producing more pronounced steps in the reduction of rope tension. The effective rope tension is governed by the transition
\begin{equation}
\{p_\text{before}, q_\text{before}\} \to \{p_\text{after}, q_\text{after}\}
\end{equation}
which typically yields larger values of $\kappa_\text{eff}$ than in Eq.~\eqref{eq:keff_angantyr}.

The fragmentation parameters are adjusted consistently with those used in Pythia.

\begin{figure}
\centering
\includegraphics[width=0.4\textwidth]{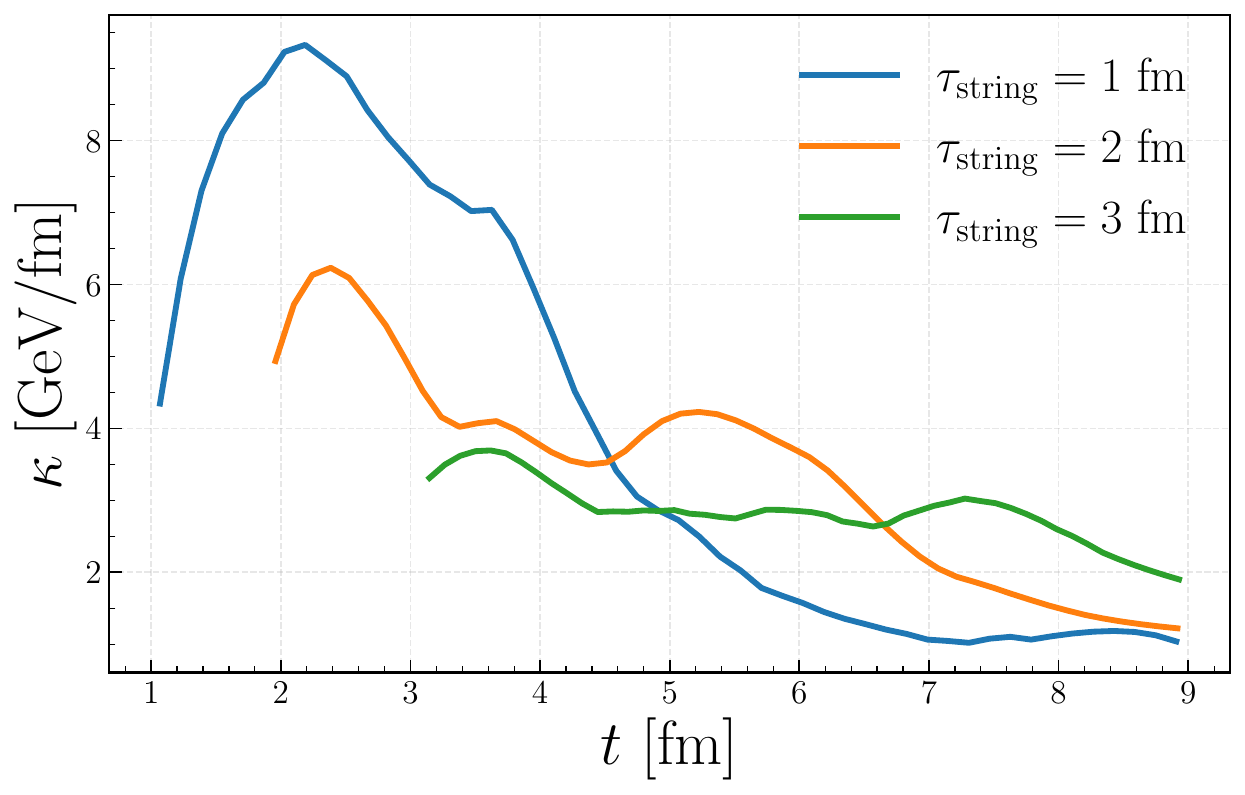}
\caption{
Time evolution of the mean effective string tension $\kappa$ in a central slice of central Pb–Pb collisions at $\sqrt{s_\text{NN}}=\SI{17.3}{GeV}$, shown for different string lifetimes $\tau_\text{string}$. Increasing the string lifetime leads to a more dilute system prior to fragmentation, reducing string overlap and consequently lowering the effective string tension.
}
\label{fig:k_eff_time}
\end{figure}

One additional parameter introduced in this work is the lifetime of a string, $\tau_\text{string}$, which controls the delay before a string fragments into hadrons. Figure~\ref{fig:k_eff_time} illustrates how $\tau_\text{string}$ affects the early-time development of the effective string tension. In this framework, strings propagate in spacetime prior to fragmentation; a longer lifetime allows them to separate further, resulting in a more dilute system and thus a smaller effective string tension.

Based on this behavior and on comparisons with hadronic observables, a single parameter set is chosen for all SMASH+Ropes results presented in the following sections. The final model parameters are listed in Table~\ref{tab:final-parameters}.

\begin{table}[htbp]
\centering
\caption{Final parameter set used in the SMASH+Ropes calculations.}
\begin{tabular}{lc}
\toprule
\textbf{Parameter} & \textbf{Value} \\
\midrule
String Stopping $\alpha$                 & $1.2$ \\
String Volume $V_{\text{string}}$        & $3.15~\mathrm{fm}^3$ \\
String Lifetime $\tau_{\text{string}}$   & $2.0~\mathrm{fm}$ \\
Time Step $dt$                           & $0.1~\mathrm{fm}$ \\
\bottomrule
\end{tabular}
\label{tab:final-parameters}
\end{table}

\begin{figure}
\centering
\includegraphics[width=\columnwidth]{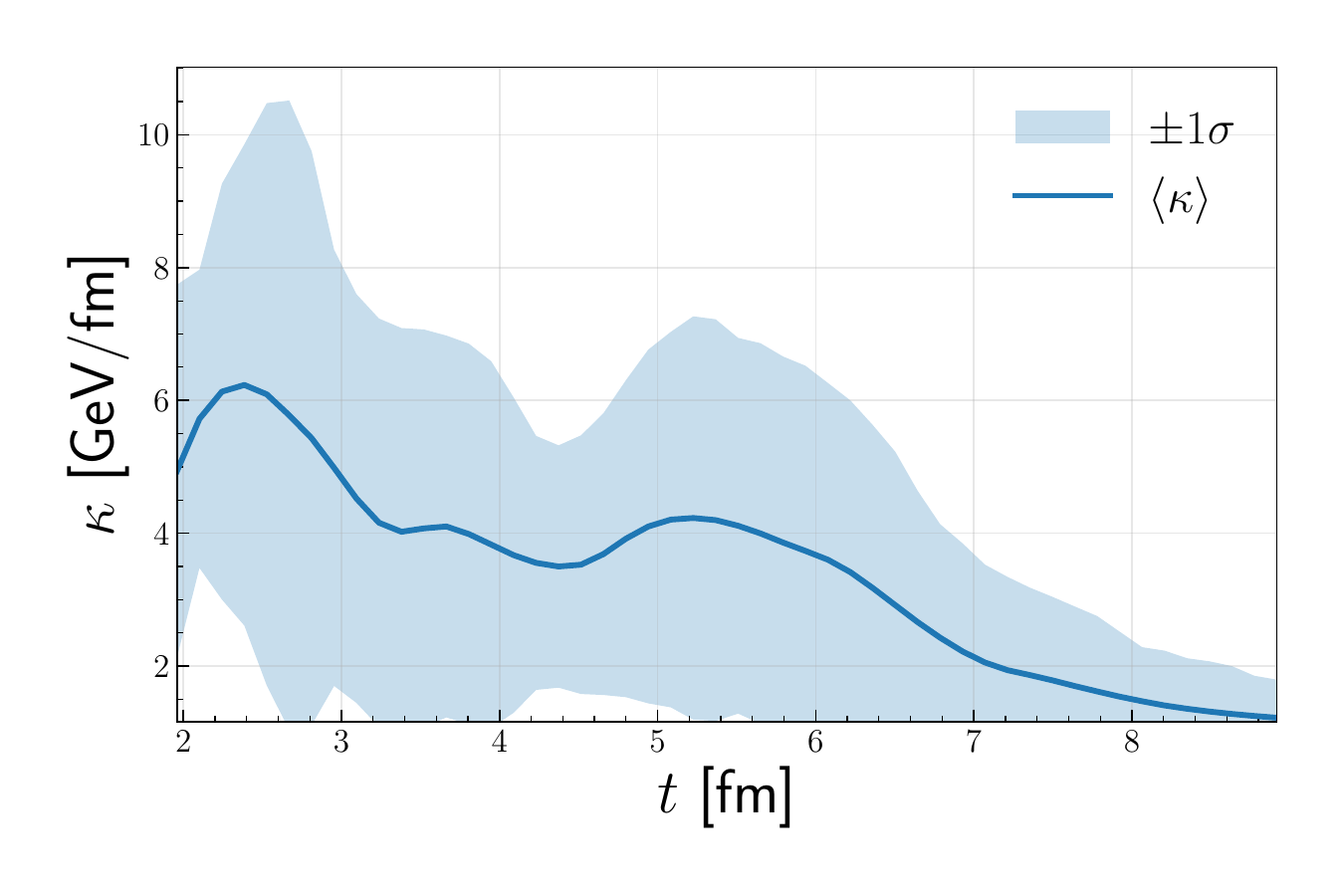}
\caption{
  Time evolution of the mean effective string tension $\kappa$ in a central slice of central Pb–Pb collisions at $\sqrt{s_\text{NN}}=\SI{17.3}{GeV}$, obtained using the parameter set given in Table~\ref{tab:final-parameters}. The shaded band indicates the event-by-event standard deviation.
}
\label{fig:k_eff_time_parm}
\end{figure}

The time evolution of the mean effective string tension for this parameter set is shown in Fig.~\ref{fig:k_eff_time_parm} and defines the baseline for the following results. Notably, the mean effective string tension exhibits two distinct peaks and remains above the vacuum value for more than the lifetime of a single string, indicating that interactions beyond the initially produced strings play a significant role in the system's dynamical evolution.

\begin{figure}
\centering
\includegraphics[width=\columnwidth]{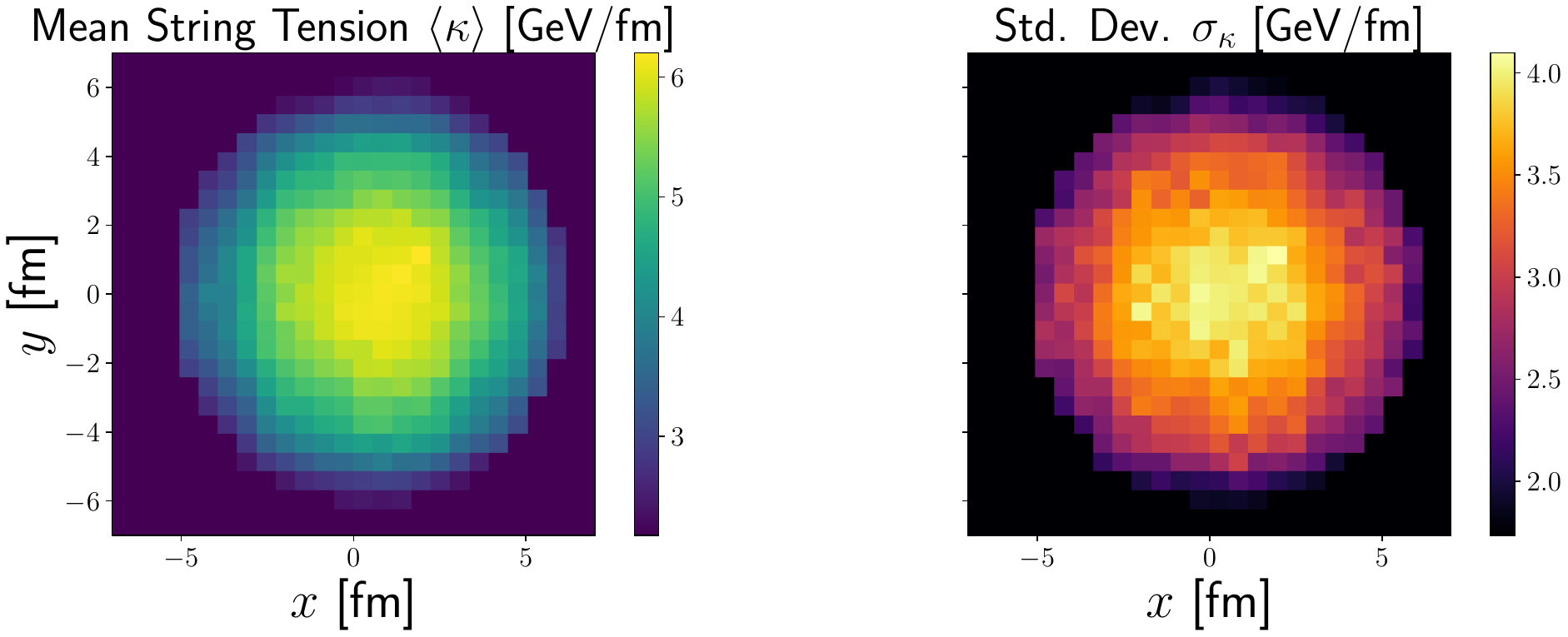}

\caption{
Effective string tension in the transverse plane at
$t \approx 3~\mathrm{fm}$ for a midrapidity slice in central
Pb--Pb collisions at
$\sqrt{s_{\mathrm{NN}}}=\SI{17.3}{GeV}$, obtained using the parameter set
of Table~\ref{tab:final-parameters}. Left: mean effective string tension
$\langle \kappa \rangle$. Right: standard deviation
$\sigma_{\kappa}$.
}
\label{fig:heat2d_kappa}
\end{figure}

To illustrate the spatial distribution of the effective string tension at an early time, Fig.~\ref{fig:heat2d_kappa} shows the effective string tension in a midrapidity slice at $t \approx 3~\mathrm{fm}$ for the same parameter set. The left panel shows the mean over many $b=0$ events. Event averaging results in a smooth transverse profile, which reflects the typical magnitude and spatial distribution of the effective string tension at this time.

The right panel shows the corresponding event-by-event standard deviation at each spatial position. Large values of $\sigma_\kappa$ indicate regions in which the local effective string tension varies strongly between events, despite the fixed centrality. Thus, the smooth mean profile masks substantial event-by-event fluctuations in the local string environment.

Together with the time evolution shown in Fig.~\ref{fig:k_eff_time_parm}, this demonstrates that the effective string tension is a strongly dynamical quantity, with substantial event-by-event fluctuations whose magnitude depends on the spatial position within the collision region.

\section{Hydrodynamic model} \label{sec:hybrid}

\begin{figure*}[t]
\centering
\includegraphics[width=\textwidth]{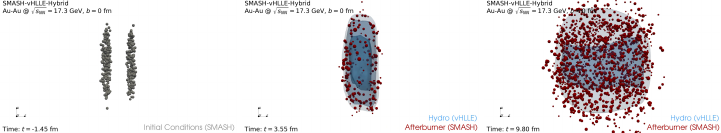}
\caption{
Three representative snapshots of the SMASH+vHLLE hybrid evolution,
showing the system at different times during a central Pb--Pb collision.
The frames illustrate the transition from the initial transport stage
to the hydrodynamic evolution as the fireball forms and expands.
}
\label{fig:hybrid_evolution}
\end{figure*}

The SMASH+vHLLE hybrid model is a framework composed of several components that describe different stages of a heavy-ion collision. The initial state is generated using SMASH, where particles from the early-stage dynamics are allowed to fluidize after crossing a specified hypersurface. These events are used to construct the energy-momentum tensor, which serves as input for the subsequent hydrodynamic evolution performed by vHLLE. A visual impression of this transition from the transport stage to the fluid-dynamic phase is shown in Fig.~\ref{fig:hybrid_evolution}.

As the system cools and becomes dilute, the SMASH hadron sampler is used to convert the fluid into hadrons on a freeze-out hypersurface. The sampled hadrons are then passed back to SMASH, which is used for the subsequent hadronic transport, including particle propagation and interactions.

The different components of the hybrid evolution are managed using the SMASH-vHLLE hybrid-handler~\cite{hybrid-handler-2.0}, while the hybrid approach itself is described in detail in Ref.~\cite{Schafer:2021csj}.

\subsection{Initial State}

In this work, the initial conditions (IC) for the hydrodynamic evolution are generated using the SMASH transport model. The early non-equilibrium stage of the collision is described microscopically, and the system is subsequently converted into a locally equilibrated, \emph{thermal} medium that serves as input for the hydrodynamic evolution.

This transition from the transport description to a thermalized fluid is achieved via a fluidization procedure, in which the energy--momentum tensor is constructed from the particle degrees of freedom. The resulting continuous fields define the initial conditions for the vHLLE hydrodynamic evolution.

Two different fluidization criteria are employed:

\begin{itemize}
\item A fixed hyperbolic time,
\begin{equation}
  \tau_0 = \frac{2 m_N}{\sqrt{s_{NN} - 4 m_N^2}} \left( R_P + R_T \right),
\end{equation}
corresponding to the nuclear passing time.

\item A dynamical fluidization criterion~\cite{Goes-Hirayama:2025nls}, where local energy density thresholds determine whether regions are converted into the thermal (hydrodynamic) phase. This leads to a core--corona separation, where dense regions are treated hydrodynamically while dilute regions remain in the transport description.
\end{itemize}
In this work, both initialization schemes are used. The energy scan results presented in Sections~\ref{res:mean_mt} and~\ref{res:ratios} are based on data generated in Ref.~\cite{Goes-Hirayama:2025nls}, which employs the dynamical fluidization criterion. In contrast, the analysis of mid-rapidity observables as a function of wounded nucleons in Section~\ref{res:mid_rap_wounded} uses the fixed-time switching criterion.

\subsection{vHLLE}

The vHLLE model~\cite{Karpenko:2013wva} is used to simulate the evolution of the hot and dense fireball formed in heavy-ion collisions. It solves the full 3+1D hydrodynamic equations for energy, momentum, and conserved charge densities, including net baryon number, electric charge, and strangeness.

The evolution in vHLLE is governed by the conservation of energy, momentum, and conserved charges:

\begin{equation}
\partial_\nu T^{\mu\nu} = 0, \quad \partial_\nu N_B^\nu = 0, \quad \partial_\nu N_Q^\nu = 0, \quad \partial_\nu N_S^\nu = 0,
\end{equation}

where $T^{\mu\nu}$ is the energy-momentum tensor, and $N_B^\nu$, $N_Q^\nu$, and $N_S^\nu$ are the net-baryon, net-charge, and net-strangeness currents, respectively. The energy-momentum tensor used in vHLLE reads:
\begin{equation}
T^{\mu\nu} = \epsilon u^\mu u^\nu - \Delta^{\mu\nu} (p + \Pi) + \pi^{\mu\nu},
\end{equation}

where $\epsilon$ is the local energy density, $p$ is the equilibrium pressure, $\Pi$ is the bulk viscous pressure, and $\pi^{\mu\nu}$ is the shear stress tensor.

For the hydrodynamic calculations performed in this work using the fixed-time fluidization criterion, we employ the same chiral mean-field equation of state as in Ref.~\cite{Schafer:2021csj}, which exhibits a smooth crossover behavior and approaches a hadron resonance gas at low energy densities. vHLLE is formulated within the second-order Israel--Stewart framework~\cite{PhysRevC.90.024912,PhysRevLett.115.132301}. For the fixed-time calculations used to study the dependence on the number of wounded nucleons presented here, however, the shear and bulk viscosities are set to zero, corresponding to the ideal-fluid limit. The energy-scan results taken from Ref.~\cite{Goes-Hirayama:2025nls} use the hydrodynamic setup described therein.

\subsection{Hadron Sampling}

The hadron sampling procedure is described in detail in~\cite{Karpenko:2015xea}; only a brief overview is provided here.

As the hydrodynamic evolution approaches low energy densities, hadrons are sampled from a hypersurface of constant energy density, typically $\varepsilon = 0.5~\mathrm{GeV/fm^3}$. This hypersurface is constructed from the local properties of the fluid, and the Cooper--Frye prescription is applied to convert fluid elements into particles. To remain consistent with the hadronic transport model, thermodynamic quantities are recalculated using a non-interacting hadron gas equation of state. Viscous corrections are included during sampling, and the resulting particles are propagated using SMASH.

\section{Results} \label{sec:results}

In this section, we present results for strange-hadron production in nucleus–nucleus collisions obtained with the approaches introduced in Sec.~\ref{sec:models}. We focus on midrapidity observables, namely particle yields, average transverse masses, and strangeness-to-pion yield ratios. Together, these observables probe both strangeness production and the transverse dynamics of the system.

The results are shown as functions of the number of wounded nucleons and the center-of-mass energy, allowing for a direct comparison between the different approaches. We focus on the low to intermediate beam-energy range, where the interplay between non-equilibrium transport dynamics and the possible emergence of locally equilibrated fluid dynamics is particularly relevant. In the string-interaction picture, this energy range is also of interest because string excitations become increasingly important for particle production.

For clarity, we first describe the results and the main trends visible in the figures. A broader interpretation of these findings and their implications for strangeness-production mechanisms is deferred to Sec.~\ref{sec:discussion}.

\subsection{Centrality Dependence}\label{res:mid_rap_wounded}

In the following, particle production is studied as a function of the number of wounded nucleons $N_\mathrm{w}$, which serves as a measure of the collision centrality. Wounded nucleons are defined as nucleons that undergo at least one inelastic interaction during the collision. 

To ensure a consistent definition across all approaches, $N_\mathrm{w}$ is estimated using the Angantyr model for all calculations, independent of the underlying particle production model. The procedure is described in Appendix~\ref{app:wounded_nucleons}.

Fig.~\ref{fig:mid_rapidity_baryons} presents the mid-rapidity yields of strange baryons for the different approaches in the context of available experimental data. SMASH consistently underproduces strange hadrons at both beam energies, including $\Lambda$, anti-$\Lambda$, and $\Xi^-$ baryons. SMASH+Ropes improves the situation: the mid-rapidity yields of $\Lambda$ and anti-$\Lambda$ baryons are reproduced reasonably well, while $\Xi^-$ baryons remain underproduced. In contrast, SMASH+vHLLE shows overall good agreement with the experimental data, although it slightly overestimates the yields at $E_{\rm lab}=158A$ GeV. 

Pythia/Angantyr underpredicts the yields of strange baryons at both collision energies, with the exception of anti-$\Lambda$, for which the experimental data are overpredicted. Including rope hadronization improves the description of the strange-baryon yields, but further increases the overprediction of the anti-$\Lambda$ yield.

\begin{figure}
\centering
\includegraphics[width=0.5\textwidth]{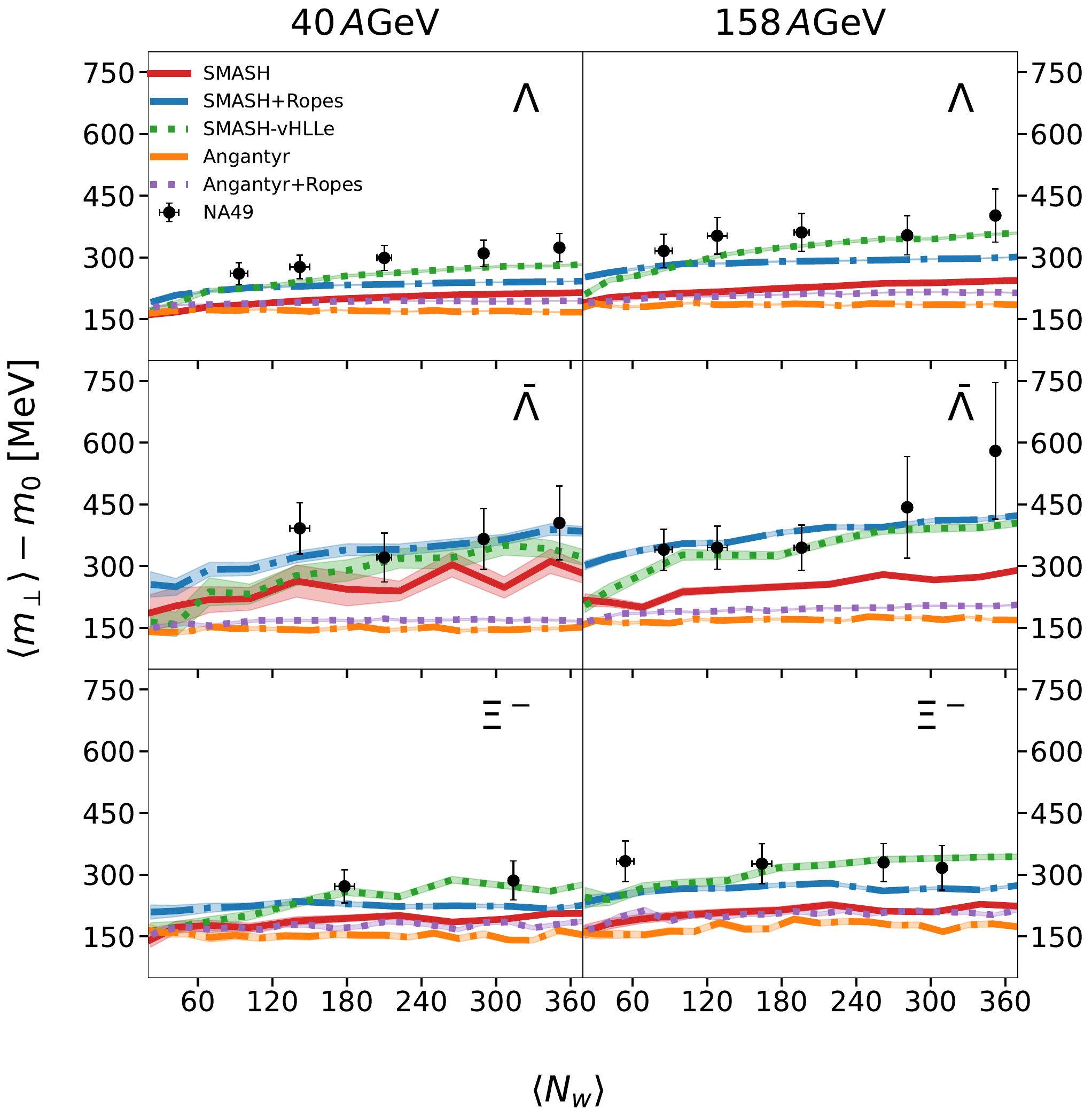}
\caption{Average transverse mass $\langle m_\perp \rangle$ minus the particle mass for $\Lambda + \Sigma^0$, $\overline{\Lambda} + \overline{\Sigma}^0$, and $\Xi^-$ at mid-rapidity as a function of the number of wounded nucleons at $E_{\rm lab}=40A$ and $158A~$GeV. Results from the models described in Sec.~\ref{sec:models} are compared with NA49 data~\cite{NA49:2009flb}.}
\label{fig:mid_transverse_mass_baryons}
\end{figure}

Fig.~\ref{fig:mid_transverse_mass_baryons} shows the average transverse mass minus the particle mass as a function of the average number of wounded nucleons for strange baryons. Similar to the mid-rapidity yields, the transverse mass is well described by SMASH+vHLLE, while SMASH underperforms. SMASH+Ropes provides an intermediate improvement: it raises the overall level compared to plain SMASH but still falls slightly below the data. Plain SMASH nevertheless captures the qualitative trend with the number of wounded nucleons, though it fails to reproduce the correct magnitude. For Angantyr, the predicted average transverse mass is consistently lower than the data and even exhibits a flat trend, in contrast to the experimental behavior.

\begin{figure}
\centering
\includegraphics[width=0.5\textwidth]{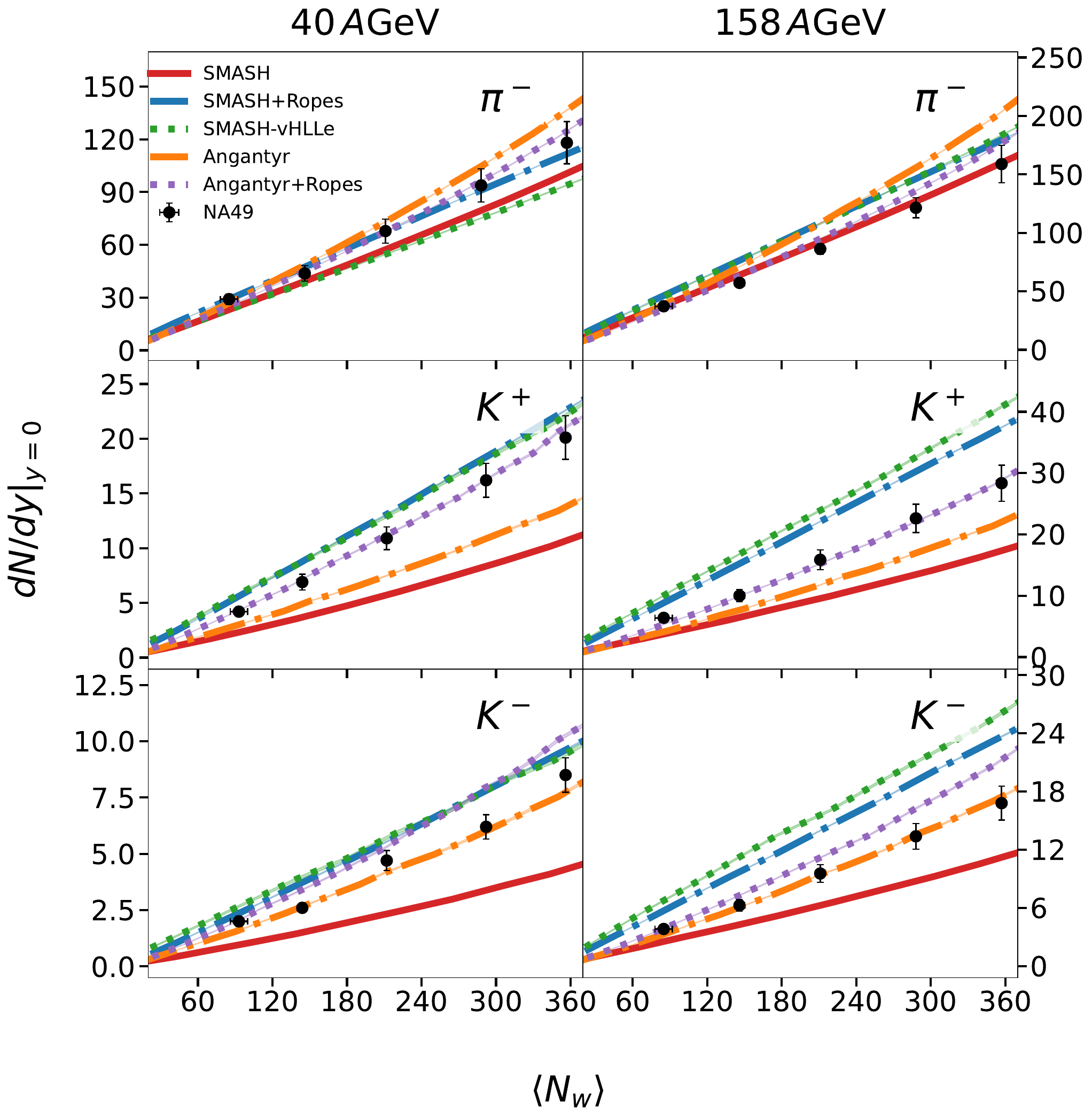}
\caption{Mid-rapidity yields of $\pi^{-}$, $K^{+}$, and $K^{-}$ as a function of the number of wounded nucleons at $E_{\rm lab}=40A$ and $158~A$GeV. Results from the models described in Sec.~\ref{sec:models} are compared with NA49 data~\cite{NA49:2012rsi}.}
\label{fig:mid_rapidity_mesons}
\end{figure}

\begin{figure}
\centering
\includegraphics[width=0.5\textwidth]{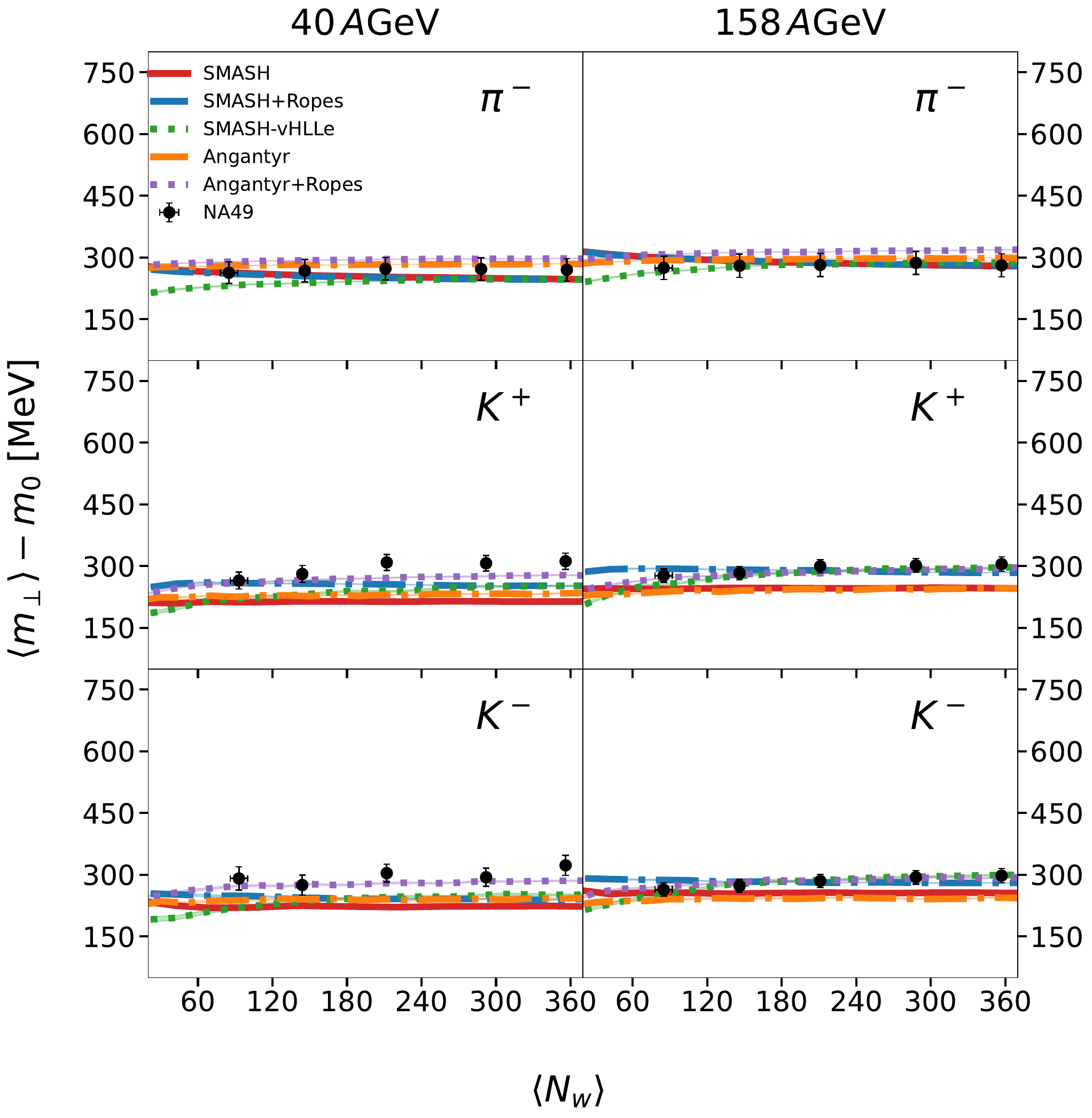}
\caption{Average transverse mass $\langle m_\perp \rangle - m_0$ for $\pi^{-}$, $K^{+}$, and $K^{-}$ at mid-rapidity as a function of the number of wounded nucleons at $E_{\rm lab}=40A$ and $158~A$GeV. Results from the models described in Sec.~\ref{sec:models} are compared with NA49 data~\cite{NA49:2012rsi}.}
\label{fig:mid_transverse_mass_mesons}
\end{figure}

Figures~\ref{fig:mid_rapidity_mesons} and~\ref{fig:mid_transverse_mass_mesons} show the mid-rapidity yields of light mesons and their corresponding average transverse masses. The pion yields are well described by all considered models. For the remaining mesons, the general trends are similar to those observed for the baryons: SMASH tends to underestimate both the yields and transverse masses, SMASH+Ropes provides a moderate improvement, and SMASH+vHLLE reproduces the overall magnitude reasonably well. Pythia/Angantyr also gives a reasonable description of the meson yields, with rope hadronization generally improving the description of strange-meson production.

\begin{figure}
\centering
\includegraphics[width=0.5\textwidth]{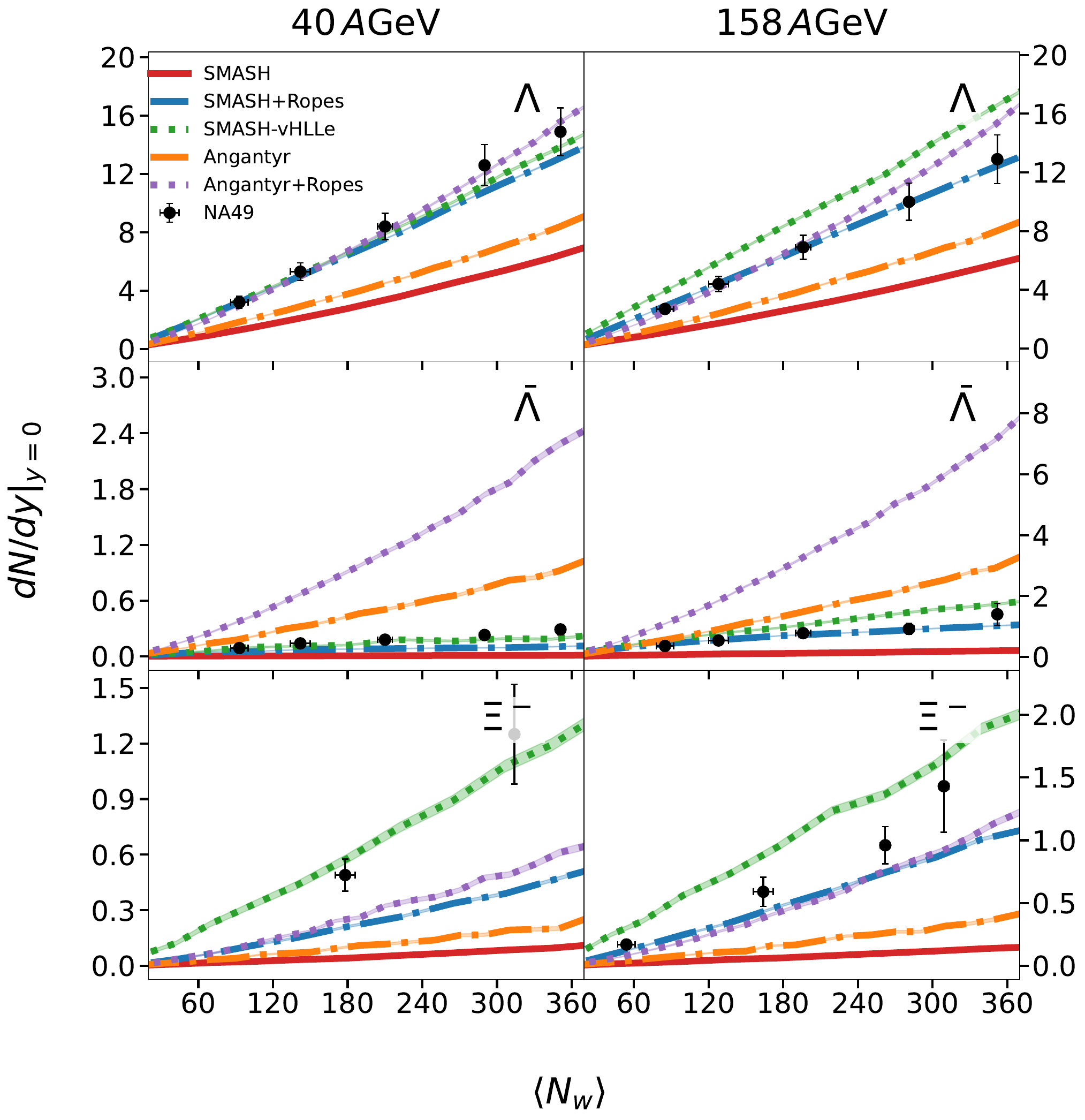}
\caption{Mid-rapidity yields of $\Lambda + \Sigma^0$, $\overline{\Lambda} + \overline{\Sigma}^0$, and $\Xi^-$ hyperons as a function of the number of wounded nucleons at $E_{\rm lab}=40A$ and $158A~$GeV. Results from the models described in Sec.~\ref{sec:models} are compared with NA49 data~\cite{NA49:2009flb}.}
\label{fig:mid_rapidity_baryons}
\end{figure}

\subsection{Energy dependence of strangeness production}\label{res:ratios}

\begin{figure}
    \centering
    \includegraphics[width=0.5\textwidth]{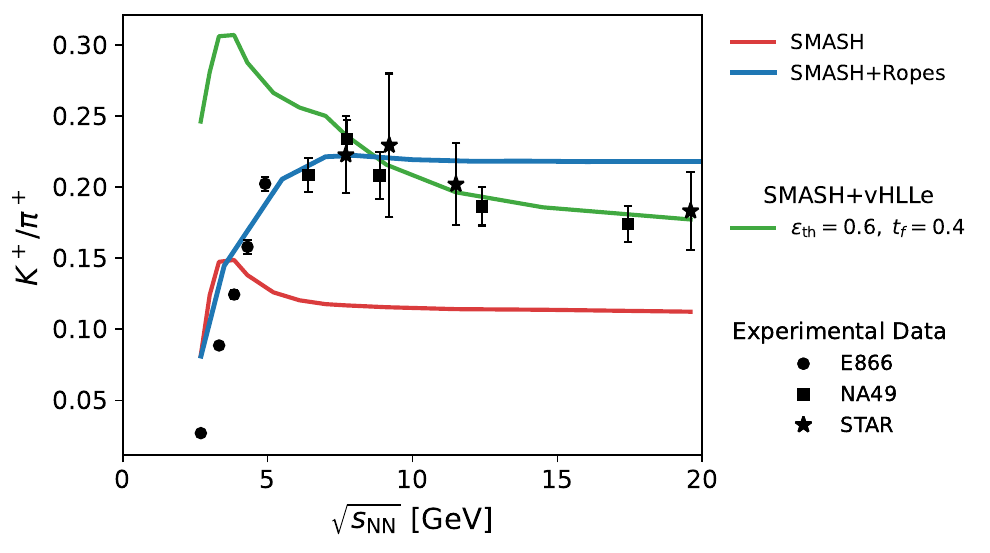}
    \caption{Midrapidity yield ratio $K^{+}/\pi^{+}$ in central Au+Au collisions as a function of $\sqrt{s_{\mathrm{NN}}}$. Results from models described in Sec. \ref{sec:models} and dynamical fluidization (taken from Ref.~\cite{Goes-Hirayama:2025nls}) are compared with data from STAR~\cite{PhysRevC.96.044904}, E866~\cite{E866:1999ktz}, and NA49~\cite{PhysRevC.66.054902}.}
    \label{fig:Kplus_over_piplus}
\end{figure}

\begin{figure}
    \centering
    \includegraphics[width=0.5\textwidth]{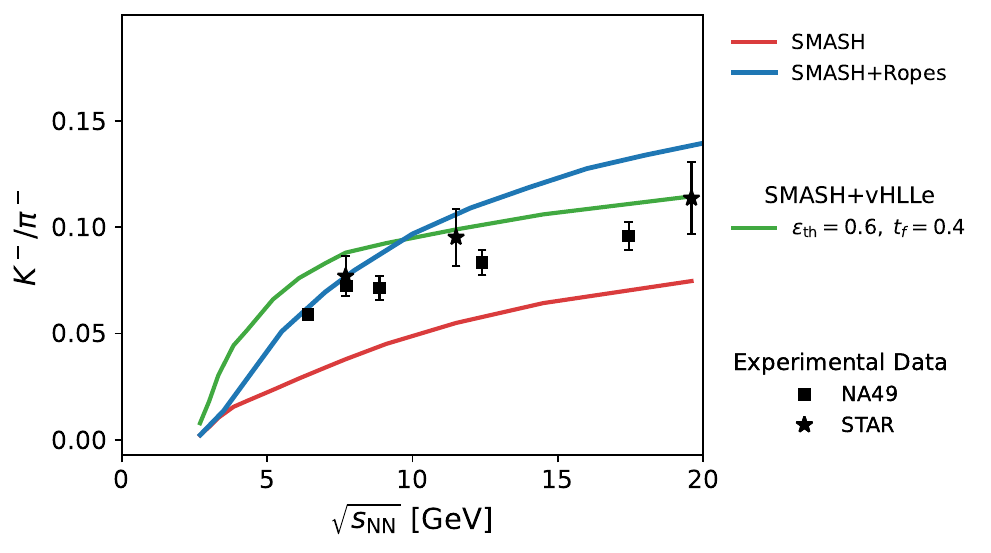}
    \caption{Midrapidity yield ratio $K^{-}/\pi^{-}$ in central Au+Au collisions as a function of $\sqrt{s_{\mathrm{NN}}}$. Results from models described in Sec. \ref{sec:models} and dynamical fluidization (taken from Ref.~\cite{Goes-Hirayama:2025nls}) are compared with data from STAR~\cite{PhysRevC.96.044904} and NA49~\cite{PhysRevC.66.054902}.}
    \label{fig:Kminus_over_piminus}
\end{figure}

\begin{figure}
    \centering
    \includegraphics[width=0.5\textwidth]{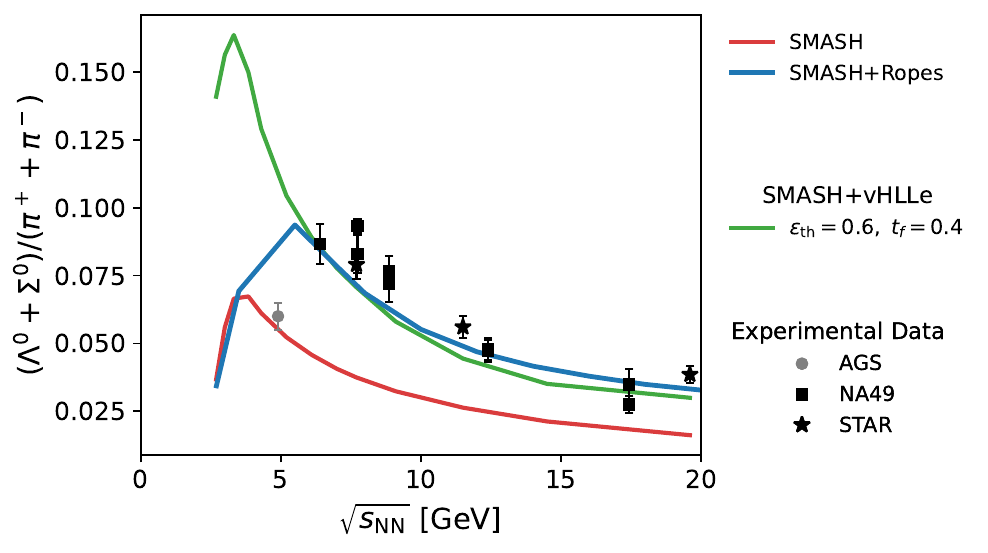}
	\caption{Midrapidity yield ratio $(\Lambda+\Sigma^{0})/\pi$ in central Au+Au collisions as a function of $\sqrt{s_{\mathrm{NN}}}$. Results from the models described in Sec.~\ref{sec:models} and the dynamical fluidization calculations of Ref.~\cite{Goes-Hirayama:2025nls} are compared with data from STAR~\cite{PhysRevC.96.044904}, NA49~\cite{PhysRevC.66.054902}, and AGS. The AGS point was digitized from Fig.~2 of Ref.~\cite{Nahrgang:2011rd}.}

    \label{fig:LambdaSigma_over_pions}
\end{figure}

The midrapidity yield ratio $K^+/\pi^+$ as a function of $\sqrt{s_\mathrm{NN}}$ in the range 
$2.5 \leq \sqrt{s_\mathrm{NN}} \leq 20~\mathrm{GeV}$ is shown in Fig.~\ref{fig:Kplus_over_piplus}.
At lower beam energies, the dynamical fluidization scenario of SMASH+vHLLE tends to
overestimate the ratio compared to the experimental data, whereas at higher energies it provides
a reasonable description of the observed trend.  
In contrast, the rope-hadronization approach reproduces the ratio well up to approximately
$\sqrt{s_\mathrm{NN}}\sim\SI{10}{GeV}$. Beyond this regime, however, the predicted ratio decreases
more slowly than in the data, leading to an overshoot.

A similar pattern is observed in Figs.~\ref{fig:Kminus_over_piminus} and
\ref{fig:LambdaSigma_over_pions}. For the latter, the high-energy overshoot of the rope approach
is noticeably smaller, while the low-energy overshoot of the hybrid model is considerably larger.

\subsection{Energy dependence of transverse dynamics}

\label{res:mean_mt}

\begin{figure}
    \centering
    \includegraphics[width=0.5\textwidth]{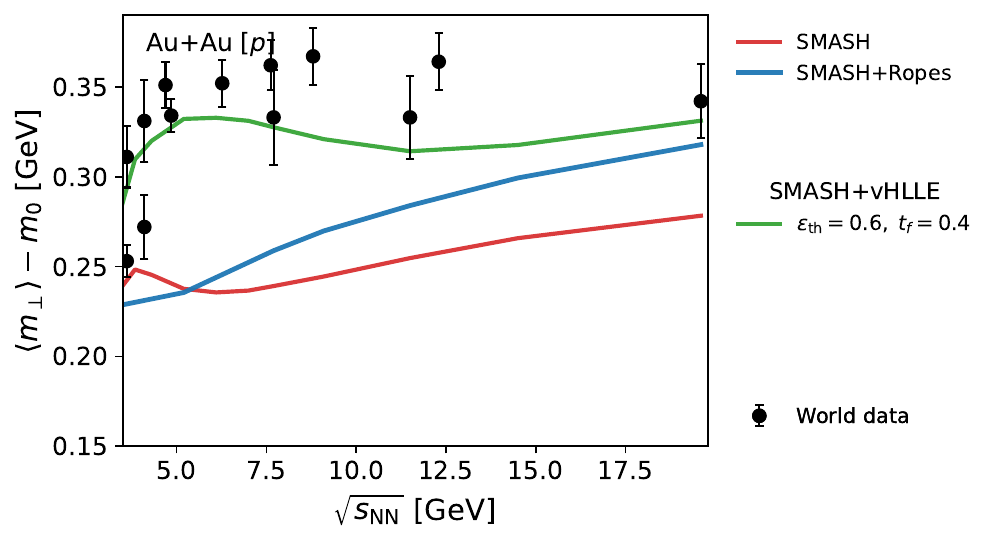}
    \caption{Average transverse mass minus rest mass, $\langle m_\perp\rangle - m_0$, at midrapidity for protons in central Pb+Pb/Au+Au collisions as a function of $\sqrt{s_{\mathrm{NN}}}$. Results from the models described in Sec.~\ref{sec:models} and dynamical fluidization (taken from Ref.~\cite{Goes-Hirayama:2025nls}) are compared with available data~\cite{STAR:2017sal,PhysRevC.69.024902}.}
    \label{fig:mtm0_proton}
\end{figure}

\begin{figure}
    \centering
    \includegraphics[width=0.5\textwidth]{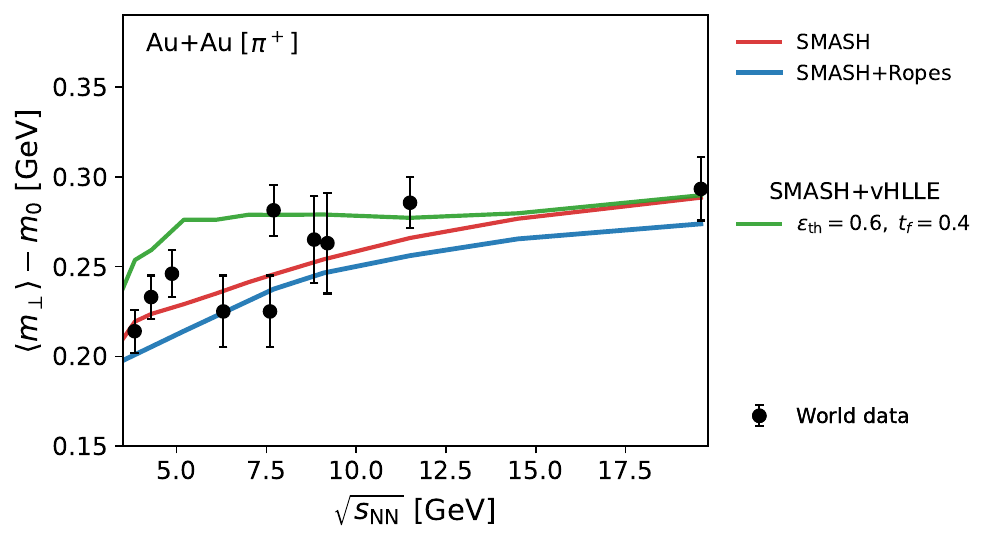}
    \caption{Average transverse mass minus rest mass, $\langle m_\perp\rangle - m_0$, at midrapidity for positively charged pions in central Pb+Pb/Au+Au collisions as a function of $\sqrt{s_{\mathrm{NN}}}$. Results from the models described in Sec.~\ref{sec:models} and dynamical fluidization (taken from Ref.~\cite{Goes-Hirayama:2025nls}) are compared with available data~\cite{STAR:2017sal,NA49:2002pzu,PhysRevC.77.024903}.}
    \label{fig:mtm0_pion}
\end{figure}

\begin{figure}
    \centering
    \includegraphics[width=0.5\textwidth]{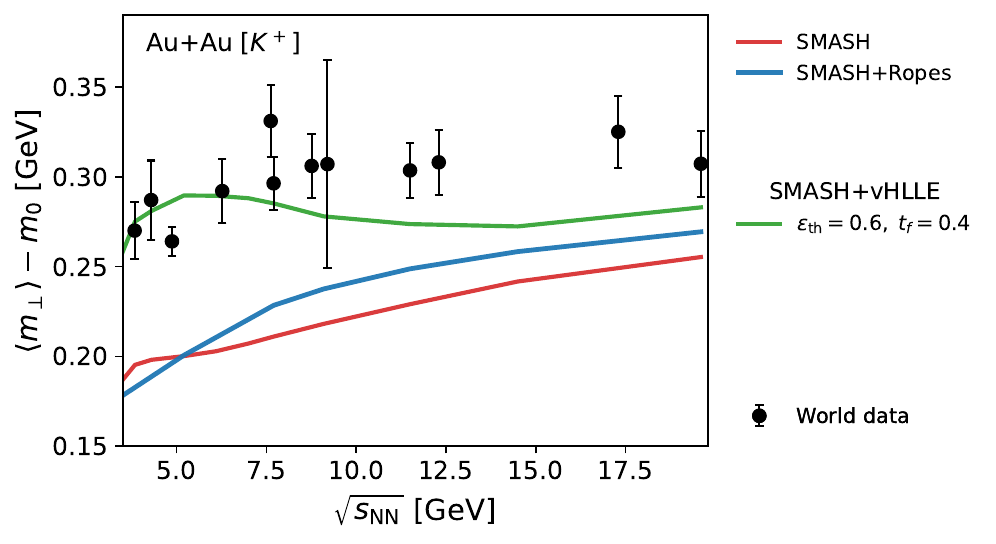}
    \caption{Average transverse mass minus rest mass, $\langle m_\perp\rangle - m_0$, at midrapidity for positively charged kaons in central Pb+Pb/Au+Au collisions as a function of $\sqrt{s_{\mathrm{NN}}}$. Results from the models described in Sec.~\ref{sec:models} and dynamical fluidization (taken from Ref.~\cite{Goes-Hirayama:2025nls}) are compared with available data~\cite{STAR:2017sal,NA49:2002pzu,PhysRevC.77.024903,E866:1999ktz}.}
    \label{fig:mtm0_kaon}
\end{figure}

The average transverse mass minus rest mass, $\langle m_\perp \rangle - m_0$, as a function of $\sqrt{s_\mathrm{NN}}$ is shown for protons, pions, and kaons in Figs.~\ref{fig:mtm0_proton}--\ref{fig:mtm0_kaon}. 
SMASH and SMASH+Ropes underperform for all particle species, while the hybrid approach describes the data well, except for charged pions.

\section{Discussion}\label{sec:discussion}

To obtain a comprehensive picture of the collision dynamics, it is essential to consider both particle yields and transverse-momentum observables. While particle yields constrain the amount of energy converted into particle production and are sensitive to entropy and flavor production, the average transverse mass $\langle m_\perp \rangle - m_0$ provides complementary information on the transverse dynamics of the system, in particular collective expansion. Considering these observables together therefore helps constrain how the available energy is distributed between particle production and collective motion.

At first glance, one might expect Angantyr and SMASH to produce similar results, as both rely on
Pythia for string fragmentation. However, the underlying mechanisms for parton-level
particle production differ substantially. SMASH employs a simplified picture of string excitation
in which two incoming hadrons exchange valence quarks, forming a single string between two
color-connected partons. Angantyr, in contrast, incorporates multiple parton interactions (MPI)
and parton showers, generating a denser and more complex initial partonic state.
Consequently, the resulting string configurations in Angantyr differ substantially from those in
SMASH.

These differences in the underlying string-production mechanisms can lead to different strange-hadron yields even when the same string-fragmentation framework is employed. Establishing an appropriate elementary strangeness-production baseline is therefore essential before attributing differences in strange-hadron production to collective effects.

A complementary picture emerges from the average transverse mass at midrapidity,
shown in Fig.~\ref{fig:mid_transverse_mass_baryons}. SMASH qualitatively reproduces the
increase of $\langle m_\perp\rangle-m_0$ with number of wounded nucleons, although it underestimates its magnitude,
while Angantyr shows a much weaker dependence. Differences in the initial string production,
together with the presence of hadronic rescattering in SMASH, may contribute to this behavior.

The SMASH+vHLLE hybrid model provides a strong overall description of both the midrapidity yields
and the average transverse mass. This behavior is consistent with earlier results obtained using
the UrQMD hybrid approach, which successfully described similar SPS data from the NA49 and NA57
Collaborations~\cite{Petersen:2009zi}. In both cases, the inclusion of a hydrodynamic phase
following an initial transport stage improves the description of particle production and
collective dynamics.

Since both the hybrid approach and the rope approach reproduce several observables reasonably
well, it is useful to compare their energy dependence directly. The $K^+/\pi^+$ ratio
(Fig.~\ref{fig:Kplus_over_piplus}), long considered a potential signal of the QCD phase
transition, highlights their respective limitations. The dynamical fluidization scenario
systematically overshoots the ratio at low $\sqrt{s_\mathrm{NN}}$, indicating that the present
implementation produces too much strangeness in this regime. A similar behavior has also been
reported for UrQMD+Hydro in Ref.~\cite{Nahrgang:2011rd}, suggesting that this may not be specific
to the present hybrid implementation.

The rope approach, by contrast, reproduces the enhancement up to
$\sqrt{s_\mathrm{NN}}\sim\SI{10}{GeV}$ but fails to capture the turnover at higher energies: the
predicted strangeness production decreases too slowly with beam energy. This points to a missing
mechanism in the current rope implementation that would moderate strangeness production at higher
energies, for example through increased pion production or a reduction of the effective string
tension. Repulsive interactions between strings, such as the string shoving mechanism
\cite{Bierlich:2020naj}, provide one possible extension. By dynamically separating overlapping
strings, such interactions could reduce the large string overlaps and thereby moderate the
resulting strangeness enhancement.

Considering also the energy dependence of the average transverse mass shown in
Figs.~\ref{fig:mtm0_proton}–\ref{fig:mtm0_kaon}, a clearer distinction between the approaches
emerges. The non-thermal models considered here do not generate sufficient collective transverse
expansion to reproduce the experimental data. Some spectral hardening is nevertheless observed
for kaons and protons in the rope approach due to the increased effective string tension, but
this effect alone is not sufficient to account for the measured average transverse masses.

The hybrid model, in contrast, develops collective expansion during the hydrodynamic stage and
reproduces the transverse-mass data reasonably well, although deviations remain for pions toward
the lower end of the energy scan. Importantly, the present comparison does not imply that
hydrodynamic evolution is the only mechanism capable of generating the required collectivity.
Rather, the non-thermal approaches studied here lack an explicit mechanism for generating
sufficiently strong collective transverse expansion. Repulsive interactions between strings,
such as string shoving, provide one possible microscopic mechanism for generating additional
transverse collectivity without requiring local thermal equilibrium.

Taken together with the strangeness-to-pion ratios, these results reveal an interesting tension.
At low collision energies, the rope approach provides a comparatively good description of
strangeness production but lacks sufficient transverse collectivity, whereas the hybrid approach
generates the required transverse expansion but tends to overpredict strangeness production.
A successful non-thermal description would therefore require an additional mechanism for
collective expansion, such as string shoving. Conversely, the behavior of the hybrid calculation
indicates that the treatment of fluidization and strangeness production toward the lower end of
the energy range requires further investigation.

The two approaches may therefore be complementary rather than mutually exclusive. Dense,
interacting string configurations could constitute a microscopic pre-equilibrium stage from
which a locally equilibrated medium subsequently emerges as the collision energy and system
density increase. Combining rope dynamics with dynamical fluidization would provide a framework
in which this possibility can be tested directly. In such a picture, the observed energy
dependence could reflect a gradual change in the relative importance of microscopic string
interactions and macroscopic fluid-like evolution, rather than a sharp transition between two
distinct dynamical regimes.

\section{Conclusion \& Outlook}\label{sec:sum}

We have compared SMASH, SMASH+Ropes, SMASH+vHLLE, and Angantyr with and without rope hadronization to study strangeness production and transverse dynamics in heavy-ion collisions at low to intermediate energies. The comparison demonstrates that both the underlying treatment of elementary particle production and the subsequent medium evolution play important roles in describing the experimental observables. In particular, the differences between SMASH and Angantyr highlight the sensitivity of strange-hadron production to the modeling of the underlying nucleon–nucleon interactions, emphasizing the need for a well-constrained elementary strangeness-production baseline before interpreting modifications in heavy-ion collisions.

Among the approaches considered here, SMASH+vHLLE provides the best overall description when both particle yields and transverse-mass observables are taken into account. The comparison with SMASH+Ropes, however, reveals complementary strengths and shortcomings of thermal and non-thermal mechanisms. At lower collision energies, the rope approach provides a comparatively good description of the strangeness-to-pion ratios, whereas the hybrid calculation tends to overpredict strangeness production. At higher energies, the rope calculation does not reproduce the observed turnover of the $K^+/\pi^+$ ratio and predicts too much strangeness relative to pion production.

The transverse-mass observables provide an important additional constraint. While the hybrid model generates sufficient collective expansion to reproduce the overall magnitude of the data, the non-thermal approaches considered here do not generate enough transverse collectivity. This does not imply that hydrodynamic evolution is the only possible origin of the observed collective behavior. Rather, it indicates that a successful non-thermal description requires an additional microscopic mechanism capable of generating collective transverse expansion.

Repulsive interactions between strings provide a natural candidate for such a mechanism. In high-multiplicity proton–proton collisions, string shoving has been shown to generate flow-like signatures and increase the average transverse momentum without requiring a thermalized medium~\cite{Bierlich:2020naj}. Extending string shoving to heavy-ion collisions could therefore address the missing transverse collectivity of the present rope implementation. At the same time, the dynamical separation of overlapping strings could reduce large rope overlaps and thereby moderate the resulting strangeness enhancement. A consistent implementation of rope formation and string shoving within a transport framework is therefore an important direction for future work.

The explicit space–time evolution available in SMASH provides a useful basis for such developments. In addition to repulsive string interactions, color reconnection offers another microscopic mechanism that can modify hadronization and baryon production. The interplay of ropes and color reconnection already provides a successful description of several strange-hadron observables in high-multiplicity proton–proton collisions~\cite{Hushnud:2023mgy}, while applications of color reconnection to heavy-ion collisions have only recently become available~\cite{Lonnblad:2023stc,Lundberg:2025ecc}. Studying these mechanisms within a common transport framework would allow their effects on strangeness production and collective dynamics to be disentangled systematically.

A complementary direction is to combine microscopic string interactions with dynamical fluidization. Dense rope configurations could then contribute to the pre-equilibrium dynamics and initial conditions, while subsequent hydrodynamic evolution generates collective expansion once local equilibration becomes applicable. Such a framework would allow thermal and non-thermal mechanisms of collectivity to be studied within the same dynamical description and could test whether their relative importance changes across the collision-energy range.

The present results therefore highlight the importance of considering strangeness production and transverse dynamics simultaneously. Strangeness enhancement alone does not uniquely distinguish between microscopic string interactions and a locally equilibrated medium, since both rope formation and hydrodynamic evolution can enhance strange-hadron production. The transverse observables provide an additional and complementary constraint, demonstrating the need for sufficiently strong collective expansion. Extending the present framework with microscopic mechanisms such as string shoving, and ultimately combining these with dynamical fluidization, provides a path toward determining when collective behavior can arise from non-equilibrium string dynamics and when a macroscopic fluid description becomes necessary.

\section{Acknowledgments}

C.~B.~Rosenkvist acknowledges fruitful discussions with R.~Góes-Hirayama and L.~Constantin, and thanks J.~Egger for providing access to the dynamical fluidization data.
Financial support from the F\&E program of GSI is gratefully acknowledged.
This research was supported in part by the cluster computing resource provided by the IT Department at the GSI Helmholtzzentrum für Schwerionenforschung, Darmstadt, Germany.

\appendix

\section{Determination of the Number of Wounded Nucleons}
\label{app:wounded_nucleons}
\begin{figure}[t]
    \centering
    \includegraphics[width=0.45\textwidth]{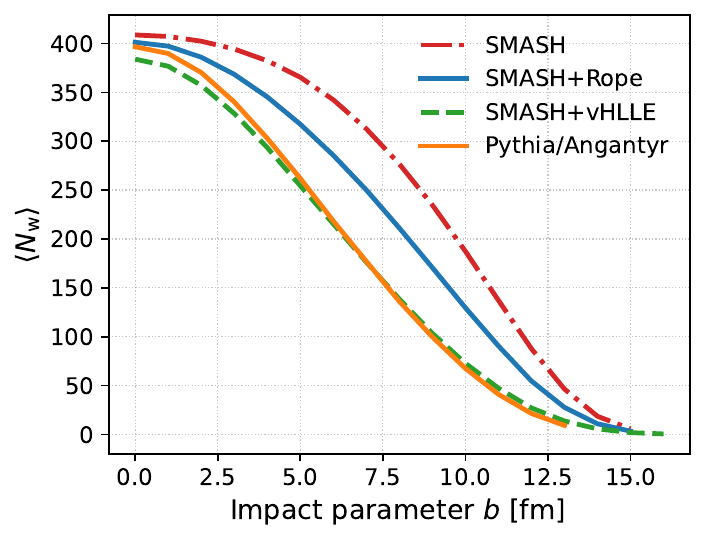}
    \caption{Mean number of participating (wounded) nucleons as a function of impact parameter for Angantyr, SMASH, SMASH+Ropes, and SMASH+vHLLE.}
    \label{fig:mean_wounded_vs_b}
\end{figure}

In transport models, the determination of the number of wounded nucleons is not unique, since late-stage hadronic interactions may excite nucleons that would be classified as spectators in a Glauber description. This can lead to an overestimation of the participant number when all interactions over the full time evolution are taken into account. One possible way to reduce this effect is to restrict the counting of wounded nucleons to early times in the evolution, as done for example in Ref.~\cite{Steinheimer:2011mp}, where good agreement with Glauber calculations was found.

In this work, we instead determine the average number of wounded nucleons as a function of impact parameter using Pythia/Angantyr. The Angantyr model provides a Glauber-based description of nucleus--nucleus collisions and is therefore well suited to define a reference relation between impact parameter and the mean number of participating nucleons.

Figure~\ref{fig:mean_wounded_vs_b} compares the mean number of wounded nucleons obtained directly from the transport evolution in SMASH, SMASH+Ropes, and SMASH+vHLLE with the Angantyr-based reference as a function of impact parameter. The transport calculations yield larger participant numbers because nucleons can undergo additional late-stage interactions that should not be counted as wounded nucleons in the Glauber sense. This effect is less pronounced in SMASH+vHLLE, where part of the evolution is described hydrodynamically, resulting in a participant count closer to the Angantyr reference. This comparison illustrates why the participant number extracted directly from the full transport evolution is not used in the present analysis. Instead, for each model variant, events are grouped according to their impact parameter, and the corresponding mean number of wounded nucleons is assigned using the Angantyr calibration.

\section{Software Versions}
\label{app:software_versions}

The versions and tags of the different software components used in this work are summarized in Table~\ref{tab:model-versions}. These reflect the state of the respective codes at the time the simulations were performed.

\begin{table}[htbp]
\centering
\caption{Versions of the physics models and tools used in this work.}
\begin{tabular}{lc}
\toprule
\textbf{Physics Software} & \textbf{Version / Tag} \\
\midrule
Pythia                  & 8.316 \cite{bierlich2022comprehensiveguidephysicsusage} \\
SMASH                    & 3.2.2 \cite{weil2025smash} \\
SMASH-vHLLE-Hybrid       & 2.0~\cite{hybrid-handler-2.0}\\
vHLLE                    & vhlle-smash-hybrid-1 \\
vHLLE parameters         & vhlle-smash-hybrid-1 \\
SMASH hadron sampler     & same as SMASH \\
\bottomrule
\end{tabular}
\label{tab:model-versions}
\end{table}

In addition, all SMASH, Angantyr, and SMASH+vHLLE hybrid outputs were analyzed using the BRASS framework~\cite{rosenkvist2025brass}, 
a general-purpose C++/Python toolkit originally developed for the binary output 
format used by SMASH.

\bibliography{ref}

\end{document}